\documentclass{INTERSPEECH2023}

\usepackage{inconsolata}
\usepackage{graphicx}
\usepackage{amsmath,amssymb}
\usepackage{multirow}
\usepackage{times}
\usepackage{latexsym}
\usepackage{amsmath}
\usepackage{algpseudocode}
\usepackage{algorithm}
\usepackage{arabtex}
\usepackage{utf8}
\usepackage{hyperref}

\hypersetup{colorlinks=true,urlcolor=blue, citecolor = blue}
\setcode{utf8}

\interspeechcameraready 

\algnewcommand\algorithmicforeach{\textbf{for each}}
\algdef{S}[FOR]{ForEach}[1]{\algorithmicforeach\ #1\ \algorithmicdo}

\title{Arabic Dysarthric Speech Recognition Using Adversarial and Signal-Based Augmentation}
\name{Massa Baali, Ibrahim Almakky, Shady Shehata\sthanks{Corresponding author.}, Fakhri Karray
}
\address{Mohamed bin Zayed University of Artificial Intelligence, Abu Dhabi, UAE}

\email{\{massa.baali,ibrahim.almakky,shady.shehata,fakhri.karray\}@mbzuai.ac.ae}

\begin{document}

\maketitle
 
\begin{abstract}
Despite major advancements in Automatic Speech Recognition (ASR), the state-of-the-art ASR systems struggle to deal with impaired speech even with high-resource languages. In Arabic, this challenge gets amplified, with added complexities in collecting data from dysarthric speakers.
In this paper, we aim to improve the performance of Arabic dysarthric automatic speech recognition through a multi-stage augmentation approach. To this effect, we first propose a signal-based approach to generate dysarthric Arabic speech from healthy Arabic speech by modifying its speed and tempo. We also propose a second stage Parallel Wave Generative (PWG) adversarial model that is trained on an English dysarthric dataset to capture language-independant dysarthric speech patterns and further augment the signal-adjusted speech samples. Furthermore, we propose a fine-tuning and text-correction strategies for Arabic Conformer at different dysarthric speech severity levels.
Our fine-tuned Conformer achieved $18\%$ Word Error Rate (WER) and $17.2\%$ Character Error Rate (CER) on synthetically generated dysarthric speech from the Arabic common voice speech dataset. This shows significant WER improvement of $81.8\%$ compared to the baseline model trained solely on healthy data.  
We perform further validation on real English dysarthric speech showing a WER improvement of $124\%$ compared to the baseline trained only on healthy English LJSpeech dataset. 
\end{abstract}

\section{Introduction}
Great progress has been achieved in Automatic Speech Recognition (ASR) systems, with remarkable performance reaching human parity \cite{amodei2016deep}. As ASR systems become more ubiquitous in daily life, individuals with impaired speech will face difficulties. However, limited work has been carried out on ASR for impaired speech recognition tasks, especially Arabic dysarthric speech. 
Dysarthria is a speech disorder caused by trauma to areas of the brain concerned with motor aspects of speech resulting in effortful, sluggish, slurred, or prosodically disordered speaking. It is often unfeasible to collect large amounts of data from dysarthric speakers due to many reasons such as low compliance and fatigue during recordings \cite{turrisi2021easycall}. This leads to data scarcity for training and testing models, even with high-resource languages.

Recently, techniques were developed to enhance the quality of automatic dysarthric speech recognition on English language \cite{hu2022exploiting,yue2022raw,geng2022speaker,joy2018improving,xie2022variational,hernandez2022cross,espana2016automatic,soleymanpour2022synthesizing,ren2017automatic}. Data augmentation using Text To Speech (TTS) was proposed by \cite{matsuzaka2022data}, where a Deep Neural Network (DNN) was trained on dysarthric speech then used to produce synthesized dysarthric speech to enhance the ASR performance with augmented data.\cite{dhanalakshmi2015intelligibility} trained an ASR using Hidden Markov Models (HMM) which synthesizes the produced text from the ASR into speech with the patient's original tone through TTS model. 
Mimicking English dysarthric speech features at the signal level, \cite{vachhani2018data} investigated data augmentation on healthy speech using temporal and speed perturbations to simulate English dysarthric speech. \cite{geng2022investigation} applied a vocal tract length perturbation  on existing dysarthric speech samples by adjusting both the speed and tempo. \cite{bhat2022improved} improved the WER by employing a two-stage data augmentation technique that involves static and dynamic augmentation of dysarthric speech data, where an end-to-end ASR model was trained to evaluate the data augmentation scheme.  Other approaches involved adversarial methods such as, \cite{jin2021adversarial} that developed a Deep Convolutional Generative Adversarial Network (DCGAN) to simulate dysarthric speech through learning a generative task to mimic the fine-grained temporal features of dysarthric speech. \cite{yu2018development} employed time-delayed neural networks and Long-Short Term Memory Networks (LSTMs) on the dysarthric speech dataset, Universal Access Speech \cite{kim2008dysarthric}, after training on two out-of-domain broadcast news and switchboard datasets. Furthermore, Meta-Learning was employed by \cite{wang2021improved} to create better base models pre-trained on large amounts of healthy speech. The meta-update of the base model repeatedly simulates adaptation to different dysarthric speakers in such a manner that optimizes the base model towards better generalization.

In this work, we aim to enhance the performance of Arabic dysarthric ASR in terms of WER using a multi-stage data augmentation method to address the lack of Arabic dysarthric speech datasets. Firstly, inspired by \cite{vachhani2018data}, we employ a signal-based approach to mimic dysarthric speech patterns and generate dysarthric speech from healthy input. Secondly, we develop an adversarial-based augmentation method using a Parallel Wave Generative (PWG) \cite{yamamoto2020parallel} model to complement the signal-based approach and capture language independent dysarthric speech patterns from English, where dysarthric speech data is available. We use the combination of augmented outputs from the signal-based and adversarial approaches to fine-tune a Conformer model to recognise Arabic dysarthric speech. Finally, we use the observed character-level confusion to build a weighted text-correction module that further improves the Conformer's performance. 

\begin{figure*}[t]
    \centering
    \centerline{\includegraphics[width=0.92\textwidth]{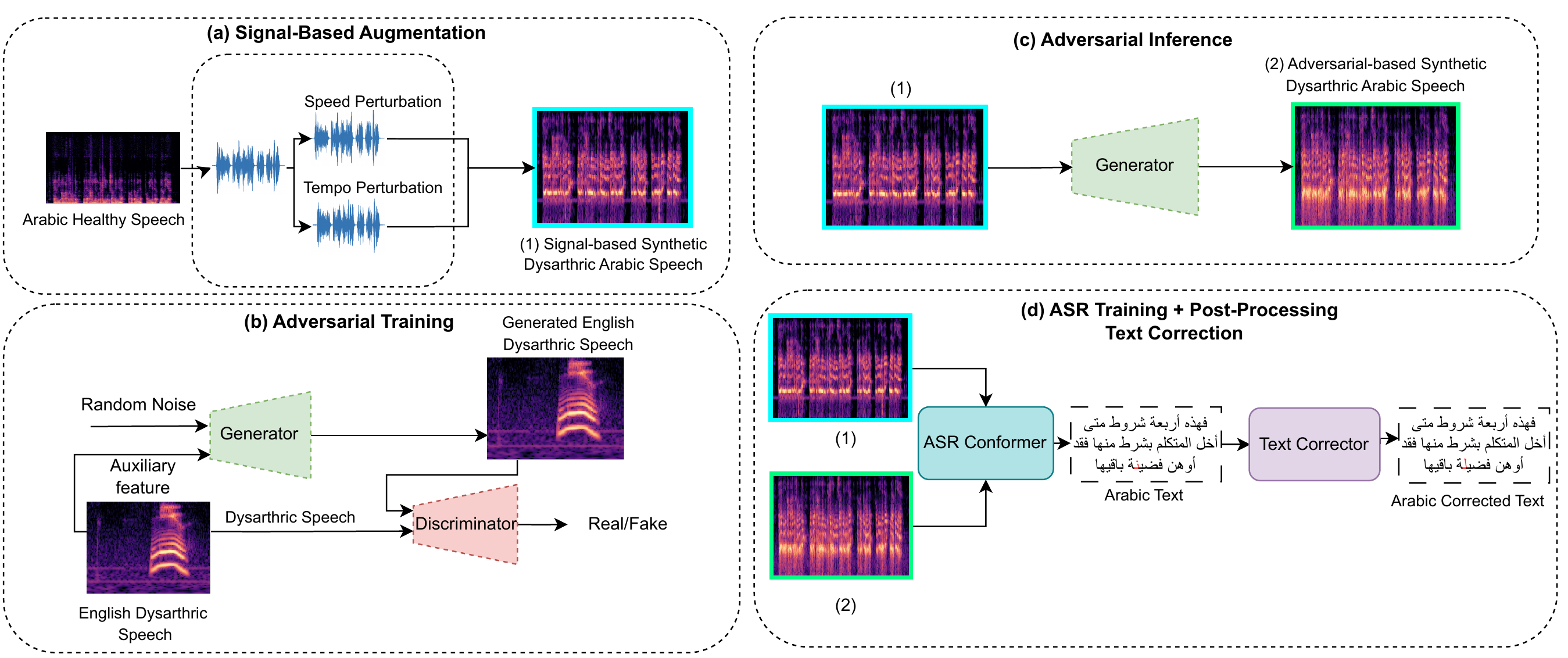}}
    \caption{Our proposed signal-based and adversarial approach used to generate Arabic dysarthric speech and fine-tune a Conformer model as well as the text correction module.}
    \label{fig:overall_pipeline}
\end{figure*}

To the best of our knowledge, this is the first study that aims to enhance Arabic dysarthric ASR by generating Arabic dysarthric speech and fine-tuning PWG model to capture language independent dysarthric speech patterns. The rest of this paper is organised as follows: Section \ref{sec:method} describes the signal-based and adversarial approach to dysarthric speech augmentation along with the Conformer model and the targeted text correction approach. Section \ref{sec:exps} then describes the experimental settings along with the datasets used for training and augmentation. We then move on to Section \ref{sec:results} that showcases our results. Finally, Section \ref{sec:lim} discusses the limitations of this work then Section \ref{sec:conclusion} concludes and provides insights into possible future work. Additionally, the code\footnote{Code and Samples: \url{https://github.com/massabaali7/AR_Dysarthric}}. implementation of this work along with the full generated dataset will be made available.

\section{Methodology}  
\label{sec:method}


Firstly, we develop a signal-based perturbation approach, where we alter the speed and tempo of available healthy speech to generate dysarthric Arabic speech which we use to improve the performance of dysarthric Arabic ASR.
Secondly, we hypothesize that dysarthric speech patterns have language-independent characteristics based on similarities between speech intelligibility tests across languages \cite{hegazi2019developing}. Therefore, we train an adversarial model to further mimic dysarthric Arabic speech features from the English language, where dysarthric speech datasets are available. Finally, we employ a text correction module supported by observed character-level confusion to enhance the ASR prediction performance. Figure \ref{fig:overall_pipeline} illustrates the main components of our approach and the rest of this section describes each of these components in detail.

\subsection{Signal-based perturbation}

\textbf{Speed perturbation.} We alter the time domain signal by re-sampling the input signal in the frequency domain by a perturbation factor $R_1$. This method changes the audio duration and the spectral shape \cite{ko2015audio}.  
Given speech signal x(t), perturbation factor $R_1$ is used to calculate output y(t) along the time axis as:  
\begin{equation}
    y(t) = x(R_1t)
    \label{eq:speed_1}
\end{equation}
In the frequency domain, signal modification is equivalent to the following: 
\begin{equation}
    X(f) \rightarrow  \frac{1}{R_1} X (\frac{1}{R_1} f)
    \label{eq:speed_2}
\end{equation}
where $X(f)$ and $\frac{1}{R_1} X (\frac{1}{R_1}f)$ represent the Fourier transform of \(x(t)\) and \(y(t)\) respectively. 

\textbf{Tempo perturbation.} We use the waveform overlap-add (WSOLA) algorithm \cite{verhelst1993overlap} to alter the duration of time domain signal by factor $R_2$, while ensuring that the pitch and spectral of the signal do not change. The time-domain input speech signal x(t) is decomposed into analysis blocks \(\bar{x}_m(r)\) that are equally spaced along the time axis by analysis hopsize. Given a perturbation factor $R_2$, the synthesis block \(\bar{y}_m(r)\) is relocated by \(H_s\) (synthesis hypothesis) based on the following equation: 
\begin{equation}
    H_s = R_2 \times H_{R_2} 
    \label{eq:tempo}
\end{equation}
An iterative process is then followed to update the analysis blocks to ensure maximal similarity between perturbed output $y(t)$ and $x(t)$.

\subsection{Adversarial Augmentation using Parallel Wave GAN}

Based on similarities between speech intelligibility tests across languages \cite{hegazi2019developing}, our hypothesis is that dysarthric speech features are language independent, we leverage an existing dysarthric English data where we trained a PWG on dysarthric English speech to generate Arabic dysarthric speech from the signal-based dysarthric mel-spectrogram. 
As shown in Figure \ref{fig:overall_pipeline}, the model consists of a generator ($G$) and discriminator ($D$). A WaveNet-based model is used as the generator, which is fed the mel-spectrogram of dysarthric speech (Auxiliary feature) along with random noise ($z$). The generator $G$ tries to learn a distribution of realistic waveforms and generate both real and fake data to deceive the discriminator $D$, which is in turn trying to recognize whether its input is real or fake examples. This leads $G$ to output speech with more realistic dysarthric features. This model is then trained using the following adversarial loss function $L$:
\begin{equation}
    L(G,D) = \mathbb{E}_{z \sim N(0,\textit{I})}[(1-D(G(z)))^2]
    \label{eq:adv_loss}
\end{equation}
The discriminator $D$ on the other hand is trained to classify whether the mel-spectrogram input is real or fake using the following criterion:
\begin{equation}
    \resizebox{.97\hsize}{!}{
    $L(G,D) =  \mathbb{E}_{x \sim p}[(1-D(x))^2] + \mathbb{E}_{z \sim N(0,\textit{I})}[D(G(z))^2]$
    }
\end{equation}
where $x$ is the target waveform and $p$ is the target waveform distribution, which is that of the signal-based dysarthric speech. 

\subsection{Conformer for Dysarthric Speech Recognition}
The data generated by signal-based perturbation and adversarial augmentation is then used to train a Conformer model \cite{guo2021recent} for automatic dysarthric speech recognition. The Conformer model comprises of a Conformer encoder and Transformer decoder, where the encoder is a multi-blocked architecture. Each block is stacked by a position-wise feed-forward network (FFN) module, a multihead self-attention (MHSA) module, a convolution module, and another FFN at the end of the block. The incorporation of positional embeddings to MHSA further improves the generalization of the model on variable lengths of input sequences. The details of training the Conformer are provided in Algorithm \ref{alg1}.

\subsection{Text Correction}
To correct errors caused by imprecise speech articulation in dysarthric speech, and to enhance the Conformer prediction, we propose an approach to utilise the character-level confusion from the Conformer model to create a text correction module. We employ a weighted Jaccard distance \cite{hancock2004jaccard}, where characters that are more likely to get mixed up in dysarthric speech would result in smaller distance. As such, the Jaccard distance $d_J$ between a predicted word $w_p$ and a ground truth word $w_g$ is given by:
\begin{equation}
    d_J(w_p, w_g) = 1 - \frac{\sum_{i}min(c_p, c_g)}{\sum_{i}max(c_p, c_g)}
    \label{eq:jaccard}
\end{equation}
where $c_p$ and $c_g$ are characters belonging to $w_p$ and $w_g$ respectively. Using a dictionary of words collected from Arabic Wikipedia \cite{wiki}, we pick a word from the dictionary $w_d$, where $d_J(w_p, w_d)$ is the smallest distance in $\{d_J(w_p, w_1), d_J(w_p, w_2), \dots, d_J(w_p, w_n)\}$, where $n$ is the total number of words in the dictionary.

\section{Experimental Setup}
\label{sec:exps}

\subsection{Datasets}
\label{subsec:data}
Due to data collection complexities and considering the low-resource nature of the language, there are no Arabic dysarthric speech datasets available. Therefore, we use four datasets in our proposed approach; two healthy Arabic datasets (Common Voice \cite{ardila2019common} and MGB-2 \cite{ali2016mgb}), one dysarthric English dataset (Torgo \cite{rudzicz2012torgo}), and one healthy English dataset (LJspeech \cite{ito2017lj}). Firstly, we use an 8-hour subset, split equally between female and male speakers, of the Common Voice healthy Arabic speech dataset \cite{ardila2019common} to generate synthetic dysarthric Arabic samples through signal-based perturbation. We resample this dataset to 16KHz to ensure consistency with MGB-2. In a similar manner, we use a 4-hour subset from the LJspeech dataset to synthetically generate 16 hours of dysarthric English speech for validation. Secondly, we use a 20-hour subset of the MGB-2 healthy Arabic speech corpus \cite{ali2016mgb} to train a Conformer model for Arabic ASR. Our MGB-2 subset is divided into 18 training hours, 1 hour for development, and 1 hour for testing.

We utilize Torgo \cite{rudzicz2012torgo}, which is an English dysarthric speech dataset, to familiarize our adversarial generator model with dysarthric speech characteristics beyond speed and tempo. We hypothesize that these characteristics are language independent based on similarities between speech intelligibility tests across languages \cite{hegazi2019developing}. Torgo contains a total of 23 hours of English speech along with transcripts from 8 speakers (5 Male and 3 Female) with Cerebral Palsy (CP) or Amyotrophic Lateral Sclerosis (ALS). Additionally, it contains speech from 7 speakers (4 males, 3 females) from a non-dysarthric control group. In this work, we only use the first cohort of the Torgo dataset with dysarthric speech. A test subset of $4$ speakers, $2$ male and $2$ female, from the Torgo dataset is also employed independently to validate the effectiveness of the signal-based perturbation approach.
  
\begin{algorithm}[t!]
\begin{algorithmic}
\Require $\text{Load pre-trained PWG model generator} $
\Statex {$\mathbf{P}_G$ and Conformer $\mathbf{C} $}
\ForEach {healthy Arabic sample $x_A(t)$ $\in X$} 
\State $y_A \leftarrow \text{sample ground truth}$
\State $\hat{x_A}(t) \leftarrow PertubateSignal(x_A(t))$
\State ${x^\prime_A}(t) \leftarrow \mathbf{P}_G({\hat x_A}(t)) $
\State $\hat{y}_A \leftarrow \mathbf{C}({\hat x_A}(t))$
\State $\Tilde{y}_A \leftarrow \mathbf{C}({x^\prime_A}(t))$
\State $Loss_1 \leftarrow L_{CTC}(\hat{y_A}(t), y_A)$
\State Backpropagation
\State $Loss_2 \leftarrow L_{CTC}(\Tilde{y}_A(t), y_A)$
\State Backpropagation

\EndFor 

\Function{PertubateSignal}{$x(t)$}
\State $R_1$, $R_2 \leftarrow$ assign values
\State ${\hat{x}}(t) \leftarrow x(R_1t)$
\State ${\hat x}(t) \leftarrow WSOLA({\hat x}(t))$
\State{$\mathbf{return}$ $\hat{x}$}
\EndFunction

 \caption{Training Arabic Dysarthric ASR Conformer}
 \label{alg1}
 
 \end{algorithmic}
\end{algorithm}

\subsection{Experiments}
\label{subsec:experiments}
There are three main steps that comprise our proposed approach namely; Signal-based perturbation, Adversarial training, and Conformer ASR training.

\textbf{Signal-based perturbation.} Signal-based augmentation is the first step used to generate dysarthric Arabic speech from the Common Voice dataset of healthy Arabic speech. Inspired by \cite{vachhani2018data}, the values for $R_1$ and $R_2$ in equations (\ref{eq:speed_2}) and (\ref{eq:tempo}) respectively, were selected based on the target severity level of the dysarthric speech. However, we carried out experimental modifications to $R_1$ and $R_2$ adapting them to Arabic speech, which are detailed in Table \ref{table:severity_wer}. We generate a total of $28k$ synthetic dysarthric Arabic utterances at different severity levels between $S_1$ and $S_4$ from $14k$ healthy Arabic utterances from the Common Voice dataset.

\textbf{Adversarial training.} We train two PWG models on subsets of the Torgo English dysarthric dataset. Each model is trained on either a female or male split of the dataset to ensure the generator's ability to output better gender-specific waveforms. We use a total of $3090$ utterances split into $2042$ and $1048$ utterances for the male and female subsets respectively.  We train both models by reducing the adversarial loss in equation (\ref{eq:adv_loss}) for $400k$ steps with a learning rate of $0.0001$. A scheduler was used to reduce the learning rate by a factor of $0.5$ every $200k$ steps.
The male and female generators for the two PWGs are then used to augment the output of the signal perturbed samples from the Common Voice dataset. As such, we generate a total of $56k$ utterances starting from the $28k$ signal perturbed samples at various severity levels. 

\textbf{Conformer ASR training.} 
We train a Conformer model using ESPNet tooklit \cite{guo2021recent} on an $18$-hour subset of the MGB-2 Arabic healthy speech dataset. We use a set of $5k$ Byte-Pair Encoding (BPE) tokens generated by the SentencePiece tokenizer \cite{kudo2018sentencepiece}. The architecture of the Conformer model has $12$ Conformer blocks in the encoder and 6 transformer blocks in the decoder. Each block has an output dimension of $512$ and a kernel size of $31$. The encoder and decoder both have 8 attention heads with 2048 feed-forward unit dimension. We use a max trainable epoch of $30$ with a Noam \cite{vaswani2017attention} learning rate scheduler set to $25k$ warmup steps with a learning rate of $0.0015$. Finally, we split the $56k$ utterances we got from the PWG model into $32k$ training, $12k$ validation, and $12k$ for testing. The training set is used to fine-tune our Conformer model for a further $30$ epochs. All experiments were run using NVIDIA A100 GPUs.

\begin{table}[t]
\centering
\caption{{\footnotesize \textit{WER(\%) comparison among different levels of our fine-tuned (Signal-based, and  combined (Adversarial+Signal based)) Conformers as well our text-correction module with the baseline Conformer model on the Common Voice female, male and combined sets. }}}
{
\scalebox{0.80}{
\begin{tabular}{lcccp{1.5cm}}
\hline
\textbf{Split} & \textbf{Baseline} & \textbf{Sig.} & \textbf{Sig.+Adv.} &  \textbf{Sig.+Adv.+ \newline Text Corr.}  \\ \hline
    \textit{Female} & 88.7  & 8.7 & 8.5 & \hspace{0.4cm} \textbf{8} \\
    \textit{Male} & 104 & 23.1 & 22 & \hspace{0.175cm} \textbf{ 20.5} \\
     \textit{Combined} & 99.8 & 20.2 & 19.7 & \hspace{0.22cm} \textbf{ 18} \\ 
 \hline
\end{tabular}
}
}
\label{table:wer_all}
\end{table}
\section{Results and Analysis}
\label{sec:results}

We validate the effectiveness of our approach in improving dysarthric speech recognition on real dysarthric English data, followed by synthetically generated Arabic speech. Firstly, we use real dysarthric English speech from the Torgo dataset \cite{rudzicz2012torgo} to assess the performance of the Conformer model \cite{guo2021recent} for English dysarthric speech recognition. We observe that fine-tuning the Conformer model using $16$ hours of  synthetically generated dysarthric speech from the LJSpeech dataset leads to a WER improvement of $124\%$ on the Torgo dataset. This is in comparison with the WER acheived using a baseline Conformer model pre-trained solely on healthy English speech from the same LJSpeech dataset. 
Secondly, we assess the ASR performance at different dysarthric speech severity levels using a signal-based perturbed Arabic test set of $3k$ utterances sampled from the total $14k$ utterances in the Common Voice dataset. We consider a Conformer model trained on MGB-2 Arabic healthy speech dataset as the \textit{baseline} model. In Table \ref{table:wer_all}, we compare its performance with that of Conformer models fine-tuned on synthetic dysarthric speech generated from the Common Voice dataset using our signal-based, and combined (signal and adversarial based) approaches. To improve the performance of the ASR and achieve better WER we trained the model by combining the dataset generated by the signal-based and the adversarial-based. Then we applied the text correction module as a post-processing step to further enhance the WER.   Our approach showed significant performance gain with a WER improvement of $81.8$\% compared with the \textit{baseline}. This gain also manifests with both male and female voices as shown in the performance on the two subsets. This performance gain demonstrates the effectivness of our augmentation approach in boosting their performance of Arabic dysarthic ASR. 

Table \ref{table:severity_wer} demonstrates our approach's ability in dealing with different dysarthric speech severities. We notice a pattern of increasing WER as the severity level increases, which validates our choice of values for $R_1$ and $R_2$ in equations (\ref{eq:speed_2}) and (\ref{eq:tempo}). This conforms with results previously achieved by \cite{vachhani2018data} on English dysarthric speech. Moreover, we observe better performance on the female subset likely due to better quality samples on the female subset compared the male one in Common Voice dataset. We further verify this argument using MOSNet \cite{mosnet} which shows better MOS for the female subset.
This hypothesis is further observed in Table \ref{table:severity_wer}, where the difference of WER values between female and male subsets starts small in $S_1$ and increases as the severity level increases to $S_4$.

Finally, we observe a high substitution rate for the Character Error, which is then significantly reduced using our augmentation steps. We can also observe that the addition of the text correction module further reduces this substitution rate. This shows that our proposed method is capable of dealing with imprecise articulation to lower the substitution rate and improve the overall CER. 

\begin{table}[t]
\centering
 \caption{{\footnotesize \textit{Breakdown of the Conformer's performance on various severities after fine-tuning on signal-based augmented data. }}}
 {
\scalebox{0.75}{
\begin{tabular}{c|cc|ccc}
\hline
\multirow{2}{*}{\textbf{Severity}}  & \multicolumn{2}{l|}{\textbf{Perturbation}} & \multicolumn{3}{c}{\textbf{WER (\%)}}                                                                        \\ \cline{2-6} 
                                    & \textbf{$R_1$}          & \textbf{$R_2$}         & \multicolumn{1}{l}{\textbf{Female}} & \multicolumn{1}{l}{\textbf{Male}} & \multicolumn{1}{l}{\textbf{Comb.}} \\ \hline
$S_1$                                  & 1.2                  & 0.8                 & 26.6                                & 28.4                              & 17.3                               \\
$S_2$                                  & 1.4                  & 0.8                 & 26.9                                & 28.8                              & 17.8                               \\
$S_3$                                  & 1.8                  & 0.4                 & 34.2                                & 42.4                              & 23.6                               \\
$S_4$                                  & 2                    & 0.4                 & 35.3                                & 47.0                              & 26.8                               \\
\multicolumn{1}{l|}{All Severities} & -                   & -                  & 28                                  & 34.4                              & 20.2                               \\ \hline
\end{tabular}
}}
\label{table:severity_wer}
\end{table}

\begin{table}[t]
\centering
\caption{{\footnotesize \textit{CER(\%) comparison among different levels of our fine-tuned (Signal-based, and combined (Adversarial+Signal based)) Conformers as well our text-correction module with the baseline Conformer. The columns ``sub.'', ``ins.'', and ``del.'' represent the number of substitution, insertion, and deletion errors,  respectively.}}}
{
\scalebox{0.78}{
\begin{tabular}{lcccc}
\hline
    \textbf{Approach}        & \textbf{Sub.} & \textbf{Ins.} & \textbf{Del.} &  \textbf{CER} [\%] \\ \hline
    \textit{Baseline}        & 58.0 & 18.0 & 23.2 & 99.2 \\
    \textit{Signal}          & 13.6 & 2.2 & 4.7 & 20.4 \\
    \textit{Signal+Adv.}     & 13.5 & 1.5 & 4.2  & 19.2 \\ 
    \textbf{\textit{Signal+Adv.+Text Corr.}} & \textbf{11.5} & \textbf{1.5} & \textbf{4.2} & \textbf{17.2} \\ 
 \hline
\end{tabular}
}}
\label{tab:cer}
\end{table} 

\section{Limitations}
\label{sec:lim}
Considering the unique complexities of dysarthric speech recognition, the signal-based part of our approach focuses on the speed and tempo variations. However, other variations contribute to those complexities such as variable recording conditions, uncontrolled body movements, and unbalanced data samples across speakers and diseases. That being said, the Parallel Wave GAN would be able to reconstruct further Dysarthric speech variations if introduced to the dataset through the signal-based component. 

\section{Conclusion}
\label{sec:conclusion}
This work proposes a novel end-to-end approach to enhance the performance of Arabic dysarthric speech recognition by augmenting Arabic dysarthric speech. Our proposed method combines signal-based and adversarial approaches to generate dysarthric Arabic speech based on healthy Arabic and dysarthric English speech datasets. These techniques are then employed to fine-tune state-of-the-art Conformer models to better recognise Arabic dysarthric speech. Using our approach, we were able to achieve $18\%$ WER on the modified Common Voice dataset with a WER improvement of $81.8\%$ over the Conformer baseline performance. In the future, we plan to reconstruct healthy speech from dysarthric speech to enable better accessibility. We also aim to apply our proposed approach on other low-resource languages.

\bibliographystyle{IEEEtran}
\bibliography{mybib}

\end{document}